\documentclass[fleqn,usenatbib,linenumbers]{mnras}% ,onecolumn

\usepackage[T1]{fontenc}
\usepackage{ae,aecompl}
\usepackage{graphicx}	% Including figure files
\usepackage{amsmath}	% Advanced maths commands
\usepackage{amssymb}	% Extra maths symbols
\usepackage{subfigure}
\usepackage{threeparttable}
\usepackage{mathrsfs}

\title[broadband emission from the KP system]
{Broadband Emission from a Kilonova Ejecta-Pulsar Wind Nebula System: Late-Time X-ray Afterglow Rebrightening of GRB~170817A}

\author[Ren \& Dai]{
J. Ren$^{1,2}$, Z. G. Dai$^{3,1}$\thanks{E-mail: daizg@ustc.edu.cn}
\\
$^{1}$School of Astronomy and Space Science, Nanjing University, Nanjing 210023, China\\
$^{2}$Key Laboratory of Modern Astronomy and Astrophysics (Nanjing University), Ministry of Education, Nanjing 210023, China\\
$^{3}$Department of Astronomy, School of Physical Sciences, University of Science and Technology of China, Hefei 230026, Anhui, China\\
}
\date{Accepted XXX. Received YYY; in original form ZZZ}

\pubyear{2021}

\begin{document}
\label{firstpage}
\pagerange{\pageref{firstpage}--\pageref{lastpage}}
\maketitle

\begin{abstract}
We study the broadband radiation behavior of a kilonova ejecta-pulsar wind nebula (PWN) system.
In this model, we jointly fit the observations of AT~2017gfo in UV-optical-IR bands
and the late-time X-ray afterglow of GRB~170817A.
Our work shows that a PWN powered by the remnant neutron star (NS) post GW170817 event
could affect the optical transient AT~2017gfo and re-brighten
the late-time X-ray afterglow of GRB~170817A.
The PWN radiation will regulate the trend of future X-ray observations
from a flattening to a steep decline until some other sources (e.g., a kilonova afterglow) become dominant.
The restricted ranges of the central NS parameters in this work are
consistent with the previous works based on the observations of AT~2017gfo only.
In addition, the new fitting result indicates that the NS wind is highly magnetized.
%Meanwhile, the PWN parameters are in a reasonable range compared with the observations of PWNe in our galaxy.
We point out that the radio and X-ray emission from a kilonova ejecta-PWN system
could be an important electromagnetic feature of binary NS mergers when a long-lived remnant NS is formed.
Therefore, observations of a kilonova ejecta-PWN system will provide important information to inferring the nature of a merger remnant.
\end{abstract}

%%%%%%%%%%%%%%%%%%%%%%%%%%%%%%%%%%%%%%%%%%%%%%%%%%%%%%%%%%%%%%%%%%%%%%%%%%%%%%%%%%%%
%%%%%%%%%%%%%%%%%%%%%%%%%%%%%%%%%%%%%%%%%%%%%%%%%%%%%%%%%%%%%%%%%%%%%%%%%%%%%%%%%%%%
\section{Introduction}
Compact binary mergers are the main sources of gravitational wave (GW) events
in the frequency range of the Advanced Laser Interferometer Gravitational-wave Observatory (LIGO)
and the Advanced Virgo GW detectors.
Among them, the mergers of binary neutron star (NS) and NS-black hole (BH)
draw a lot of attention since they are also potential sources of electromagnetic radiation (EM).
The first GW signal from a NS–NS merger was detected by the advanced LIGO
and Virgo detectors on 2017 August 17 12:41:04 UT \citep{Abbott_BP-2017-Abbott_R-ApJ.848L.13A}.

The nature of the merger remnant of the GW170817 event has been debated so far.
Based on the observational fact that the prompt EM signal is delayed about 1.7~seconds after the GW signal
as well as the sufficiently heavy gravitational mass of the merger remnant\citep{Abbott_BP-2017-Abbott_R-ApJ.848L.12A,Goldstein_A-2017-Veres_P-ApJ.848L.14G,
Savchenko_V-2017-Ferrigno_C-ApJ.848L.15S,Zhang_BB-2018-Zhang_B-NatCo.9.447Z},
a short-lived NS remnant seems likely (e.g., \citealp{Metzger_BD-2018-Thompson_TA-ApJ.856.101M}).
%A lot of radioactive heavy elements appear in the outflowing material owing to the $r$-process nucleosynthesis
%(\citealp{Burbidge_EM-1957-Burbidge_GR-RvMP...29..547B,Cameron_AGW-1957PASP...69..201C},
%see \citealp{Cowan_JJ-2021-Sneden_C-RvMP...93a5002C} for a recent review).
%Correspondingly, a kilonova powered by the radioactive decay of heavy elements occurs
%(\citealp{Li_LX-1998-Paczynski_B-ApJ.507L.59L}; also see \citealp{Metzger_BD-2019-LRR....23....1M} for a review),
%e.g., AT~2017gfo (\citealp{Coulter_DA-2017-Foley_RJ-Sci.358.1556C}).
%In different post-merger stages, the outflow has different properties to affect the outcomes of synthesized elements
%(e.g., \citealp{Nakar_E-2020PhR...886....1N} for further information).
%Thus, the ``red'', ``blue'', and even ``purple'' components emerge in the observations of kilonova
%(e.g., \citealp{Villar_VA-2017-Guillochon_J-ApJ.851L.21V, Zhu_JP-2020-Yang_YP-ApJ.897.20Z}).
%Based on this multi-component prescription,
%in plenty of works, some estimates on the ejecta properties by modelling and fitting AT~2017gfo were presented
%(e.g., \citealp{Kasen_D-2017-Metzger_B-Natur.551.80K, Villar_VA-2017-Guillochon_J-ApJ.851L.21V,
%Cowperthwaite_PS-2017-Berger_E-ApJ.848L.17C, Waxman_E-2018-Ofek_EO-MNRAS.481.3423W}).
Although the time-lag between the GW and EM signals point to a short-lived NS remnant,
recent works on numerical simulation and analytical calculations
have attributed this phenomenon to the effect of jet propagation
(e.g., \citealp{Beniamini_P-2020-Duran_RB-ApJ.895L.33B,
Hamidani_H-2020-Kiuchi_K-MNRAS.491.3192H,Hamidani_H-2021-Ioka_K-MNRAS.500..627H, Lyutikov_M-2020-MNRAS.491.483L,Lazzati_D-2019-Perna_R-ApJ.881.89L,
Lazzati_D-2020-Ciolfi_R-ApJ.898.59L,Ren_J-2020-Lin_DB-ApJ.901L..26R,
Pavan_A-2021-Ciolfi_R-MNRAS.tmp.1597P}, and reference therein).
Therefore, the observations are not inconsistent with the view that a long-lived NS was formed after the merger.
\cite{Yu_YW-2013-Zhang_B-ApJ.776L..40Y}
first suggested the possibility of the existence of a remnant-NS-powered kilonova, namely a mergernova.
Works have been done under this picture to comprehend the behavior of AT~2017gfo
(e.g., \citealp{Matsumoto_T-2018-Ioka_K-ApJ.861.55M,
Li_SZ-2018-Liu_LD-ApJ.861L.12L,Yu_YW-2018-Liu_LD-ApJ.861..114Y,
Ren_J-2019-Lin_DB-ApJ.885.60R}).

The observations of the GRB~170817A afterglow have lasted about four years
(e.g., \citealp{Makhathini_S-2021-Mooley_KP-ApJ.922.154M,Balasubramanian_A-2021-Corsi_A-ApJ.914L..20B,
Hajela_A-2021-Margutti_R-arXiv210402070H,Troja_E-2021-O'Connor_B-2021MNRAS.tmp.3213T}).
The popular interpretation of the origin of GRB~170817A afterglow nowadays is an off-axis observed structured jet
(e.g., \citealp{Ren_J-2020-Lin_DB-ApJ.901L..26R}).
Some other scenarios have been proposed, say,
a refreshed jet \citep{Lamb_GP-2020-Levan_AJ-ApJ.899..105L}
or an $e^\pm$-wind-injected jet \citep{Li_L-2021-Dai_ZG-ApJ.918.52L}.
About 3.4~years after the trigger of the GW170817 event, the Chandra X-ray Observatory detected
an X-ray source located at the position of GRB~170817A/AT~2017gfo
\citep{Hajela_A-2021-Margutti_R-arXiv210402070H,Hajela_A-2021-Margutti_R-GCN.29375....1H}.
But the unabsorbed X-ray flux derived from the observed data exceeds the conjectural extension
based on the off-axis structured jet model.
Since the flux excess is obvious at the X-ray band but undetected at the radio band,
an additional source instead of the GRB jet external shock to explain the data is naturally needed.
At present, the kilonova afterglow gets increasing attention.
It has been supposed to be the cause of GRB 170817A rebrightening
\citep{Hajela_A-2021-Margutti_R-arXiv210402070H,Nedora_V-2021-Radice_D-MNRAS.506.5908N}.
%The kilonova afterglow is a long-lasting, slowly rebrightening nonthermal emitter
%that originates from the interaction between the fast-tail of merger dynamic ejecta
%and the ambient material of binaries.
However, some other explanations, e.g., the $e^\pm$-wind-injected jet
\citep{Geng_JJ-2018-Dai_ZG-ApJ.856L..33G,Li_L-2021-Dai_ZG-ApJ.918.52L},
the fall-back accretion of central BH
\citep{Ishizaki_W-2021-Ioka_K-ApJ.916L.13I,Ishizaki_W-2021-Kiuchi_K-ApJ.922.185I,
Metzger_BD-2021-Fernandez_R-ApJ.916L.3M},
and maybe the counter-jet
\citep{Li_LB-2019-Geng_JJ-ApJ.880...39L,Troja_E-2021-O'Connor_B-2021MNRAS.tmp.3213T} are possible scenarios.

In this paper, we introduce a different scenario in which a pulsar wind nebula (PWN) is powered by the remnant NS
of the GW170817 event to explain the optical transient AT~2017gfo and
the late time rebrightening X-ray afterglow of GRB~170817A.
This paper is organized as follows.
In Section~{\ref{basical model}} we present our numerical model.
In Section~{\ref{fitting method}} we describe the fitting process in detail.
We discuss the significance of the fitting results in Section~{\ref{discussion}}.
We summarize our conclusions in Section~{\ref{Conclusion}}.
Throughout this work, we use the notation $Q=10^xQ_x$ in the c.g.s. unit unless noted otherwise,
and $D_L=40$~Mpc is the luminosity distance of AT~2017gfo/GRB~170817A.

%%%%%%%%%%%%%%%%%%%%%%%%%%%%%%%%%%%%%%%%%%%%%%%%%%%%%%%%%%%%%%%%%%%%%%%%%%%%%%%%%%%%%%%%%%%%%%%c%%%%%%%%%%%%%%%%%
%%%%%%%%%%%%%%%%%%%%%%%%%%%%%%%%%%%%%%%%%%%%%%%%%%%%%%%%%%%%%%%%%%%%%%%%%%%%%%%%%%%%%%%%%%%%%%%%%%%%%%%%%%%%%%%%
\section{The model}\label{basical model}
Similar to the Crab supernova remnant,
a PWN could be formed if a pulsar is left at the center of a merger remnant.
In the previous works of an energy-injected kilonova,
the released magnetic dipole (MD) radiation from the central NS is considered as an injected thermalization energy source
(e.g., \citealp{Yu_YW-2013-Zhang_B-ApJ.776L..40Y,Yu_YW-2018-Liu_LD-ApJ.861..114Y,
Kasen_D-2015-Fernandez_R-MNRAS.450.1777K}).
A PWN is formed by the interaction between the pulsar wind and ejecta after the merger
(e.g., \citealp{Kotera_K-2013-Phinney_ES-MNRAS.432.3228K,Murase_K-Toomey_MW-2018ApJ.854...60M}).
Based on those works, \cite{Ren_J-2019-Lin_DB-ApJ.885.60R}
considered a PWN embedded in the merger ejecta to explain the behavior of AT~2017gfo.
\cite{Wu_GL-2021-Yu_YY-A&A.654A.124W}
adopted the same picture in their study of GRB~160821B afterglow with slightly different modeling.

The spin-down energy of the central NS first powers a PWN.
Then, nonthermal photons radiated from the PWN will cross the ejecta shell and some photons
are absorbed to thermalize the ejecta material while
the other photons escaping from the PWN could impact the observations.
The observed flux thus consists of the emission $F_{\nu}^b$ from the ejecta
and the leaked part $F_{\nu}^{\rm leak}$ from the PWN, i.e.,
\begin{equation}
F_{\nu}^{\rm tot}=F_{\nu}^b+F_{\nu}^{\rm leak}.
\end{equation}
The estimates of $F_{\nu}^b$ and $F_{\nu}^{\rm leak}$ are given in Sections~\ref{Sec:ejecta} and \ref{Sec:PWN}, respectively.

After the ejecta becomes transparent to X-ray photons, an additional observable signal from the PWN may appear.
Both the early kilonova behavior and the late X-ray band observations may be affected.
The kilonova ejecta-PWN system was proposed in the previous work
\citep{Ren_J-2019-Lin_DB-ApJ.885.60R}.
We present some key formulas and new refinements below.

\subsection{Emission from the ejecta}\label{Sec:ejecta}
The dynamics and emission of the quasi-isotropic ejecta are implemented based on
a simplified radiation transfer model given by
\cite{Kasen_D-2010-Bildsten_L-ApJ.717..245K} and \cite{Metzger_BD-2019-LRR....23....1M}.
The merger ejecta expanding homologously is divided into $N (\gg1)$ layers
with different expansion velocities $v_i$, where $v_1=v_{\rm min}$ and $v_N=v_{\rm max}$.
The location of the $i$th layer at time $t$ is $R_i=v_i t$,
and the mass of the $i$th layer is
${m_i} = \int_{{R_{i}}}^{{R_{i+1}}} {4\pi {r^2}{\rho _{{\rm{ej}}}}(r,t)dr}$
with (\citealp{Nagakura_H-2014-Hotokezaka_K-ApJ.784L..28N})
\begin{equation}\label{Eq.rho}
\rho_{\rm ej}(r,t)={\frac{(\delta-3)M_{\rm ej}}{4\pi
R_{\max}^3}}\left[\left({R_{\min}\over
R_{\max}}\right)^{3-\delta}-1\right]^{-1}\left({r\over
R_{\max}}\right)^{-\delta},
\end{equation}
where $M_{\rm ej}$ is the total mass of the ejecta.
Evolution of the initial energy $E_{i}$ for the $i$th layer can be described by
\begin{equation}\label{dE}
{dE_{i}\over dt}=(1-e^{-\Delta \tau_{i}})e^{-\tau_{i}}
\xi L_{\rm md}+m_{i}\dot{q}_{\rm r}\eta_{{\rm th}}-{E_{i}\over
R_{i}}{dR_{i}\over dt}-L_{i}\;.
\end{equation}
Here, in each layer, the first term on the right-hand side describes the absorption of PWN emission,
the second term is the radioactive heating rate,
the third term is the adiabatic cooling rate,
and the last term is the radiation cooling rate, respectively.
The details about the parameters are presented as follows.
\begin{enumerate}
\item
The power of the pulsar wind $L_{\rm md}$ from the NS can be estimated by MD radiation, i.e.,
\begin{equation}
L_{\rm md}(t)=L_{\rm md,0}\left(1+{t\over t_{\rm sd}}\right)^{-\alpha}\label{Lsdt}
\end{equation}
with
\begin{equation}\label{luminosity}
L_{\rm md,0}={B_{\rm p}^2R^6\Omega_0^4\over{6c^3}}= 9.6\times10^{42}R_6^{6}B_{\rm p,12}^{2}P_{0,-3}^{-4}
~\rm erg \cdot s^{-1},
\end{equation}
where $\Omega_0$, $R$, $B_p$, $P_0$ and $c$ are the initial angular frequency, \textbf{the} radius,
the surface polar magnetic field, the initial spin period of the NS, and the speed of light, respectively.
Based on the fitting results of \cite{Yu_YW-2018-Liu_LD-ApJ.861..114Y} and \cite{Ren_J-2019-Lin_DB-ApJ.885.60R},
the spin-down timescale $t_{\rm sd}$ can be taken to be
\begin{equation}\label{tsdgw}
t_{\rm sd}=
{\frac{5c^5}{128GI\epsilon^2\Omega_0^4}}=9.1\times10^{5}\epsilon_{-4}^{-2}I_{45}^{-1}P_{0,-3}^4~\rm s,
\end{equation}
with $\alpha=1$ for the GW-dominated spin-down loss regime,
where $G$ is the gravitational constant, $I$ is the stellar moment of inertia,
and $\epsilon$ is the NS ellipticity.
$\xi$ describes the fraction of $L_{\rm md}$ that can be absorbed by the ejecta.
In addition, $\tau_{i}$ is the optical depth from the innermost layer to the $i$th layer
and can be described by $\tau_{i}=\sum_1^{i-1} {\Delta {\tau _i}}$
with $\Delta \tau_i=\int_{R_{i}}^{R_{i+1}} {\kappa \rho(r) dr}$.
%$\tau_{\rm tot}=\sum_1^{N-1} {\Delta {\tau _i}}=\int_{R_{\min}}^{R_{\max}} {\kappa \rho(r) dr}$ is the total optical depth of the whole ejecta in the line of sight.

\item
The radioactive power per unit mass $\dot{q}_{\rm r}$
and the thermalization efficiency of the radioactive power $\eta_{\rm th}$ can be estimated by
(\citealp{Korobkin_O-2012-Rosswog_S-MNRAS.426.1940K}; \citealp{Barnes_J-2016-Kasen_D-ApJ.829..110B};
\citealp{Metzger_BD-2019-LRR....23....1M})
\begin{equation}\label{eq:radioa}
\dot{q}_{\rm r}=4\times10^{18}\left[{1\over2}-{1\over \pi}{\rm
arctan}\left({{t-t_0}\over \sigma}\right)\right]^{1.3}\rm erg{\cdot}s^{-1}{\cdot}g^{-1}
\end{equation}
and
\begin{equation}\label{eq:effic}
\eta_{\rm th}=0.36\left[\exp\left(-0.56 t_{\rm day}\right)+{\ln (1+0.34t_{\rm
day}^{0.74})\over 0.34t_{\rm day}^{0.74}}\right],
\end{equation}
respectively. Here, $t_0=1.3$~s, $\sigma=0.11$~s, and $t_{\rm day}=t/1~\rm day$.

\item
The luminosity of the $i$th layer $L_i$ is estimated by
\begin{equation}
 L_i= \frac{E_i}{\max \left\{ t_{\rm d}^i , t_{\rm lc}^i \right\} },
\end{equation}
where the diffusion timescale $t_{\rm d}^i$ of photons reads
\begin{equation}
t_{\rm d}^i \simeq \frac{\kappa}{\beta R_i c}{\sum\limits_{j = i}^{N-1} {m_j}},\label{t_d}
\end{equation}
where $t_{\rm lc}^i=R_i/c$ is the light crossing time.
Here $\beta\simeq13.7$ is adopted (\citealp{Arnett_WD-1982ApJ.253..785A}).
\end{enumerate}

The total bolometric luminosity $L_{\rm bol}$ of the ejecta
is estimated by
\begin{equation}
L_{\rm bol}={\sum\limits_{i=1}^{N-1}}L_{i}.
\end{equation}
We assume that a blackbody spectrum of the ejecta is emitted from the photosphere at $R_{\rm ph}$
and the effective temperature $T_{\rm eff}$ is described as
(\citealp{Yu_YW-2013-Zhang_B-ApJ.776L..40Y}; \citealp{Xiao_D-2017-Liu_LD-ApJ.850L..41X};
\citealp{Li_SZ-2018-Liu_LD-ApJ.861L.12L})
\begin{equation}
T_{\rm eff}=\left({L_{\rm bol}\over4\pi\sigma_{\rm SB} R_{\rm ph}^2}\right)^{1/4},
\end{equation}
where $\sigma_{\rm SB}$ is the Stephan-Boltzmann constant.
The photosphere radius $R_{\rm ph}$ is estimated
by setting $\tau_{\rm ph}=\int_{R_{\rm ph}}^{R_{\max}}\rho(r)dr=1$ since $\tau_{\rm tot}>1$.
If $\tau_{\rm tot}\leq 1$, we fix $R_{\rm ph}$ to $R_{\min}$.
The flux density at frequency $\nu$ from the ejecta is given by
 \begin{equation}
F_{\nu}^b={2\pi h\nu^3\over c^2}{1\over \exp(h\nu/kT_{\rm eff})-1}{R_{\rm ph}^2\over D_L^2},
\end{equation}
where $h$ is the Planck constant and $k$ is the Boltzmann constant.

%%%%%%%%%%%%%%%%%%%%%%%%%%%%%%%%%%%%%%%%%%%%%%%%%%%%%%%%%%%%%%%%%%%%%%%%%%%%%%%%%%%%%%%%
%%%%%%%%%%%%%%%%%%%%%%%%%%%%%%%%%%%%%%%%%%%%%%%%%%%%%%%%%%%%%%%%%%%%%%%%%%%%%%%%%%%%%%%%
\subsection{Emission from the PWN}\label{Sec:PWN}
At the interface between the shocked and unshocked pulsar wind (``termination shock''),
electrons and positrons (leptons, hereafter) carried in the cold pulsar wind are accelerated,
and the magnetic field is amplified.
The accelerated leptons and the amplified magnetic field fill the PWN out to the radius $R_{\rm PWN}$.
Assuming $\epsilon_B$ to describe the fraction of the magnetic energy density of the total energy density behind the shock,
the magnetic energy density $U_B^{\rm PWN}$ in the PWN can be parameterized as (\citealp{Tanaka_SJ-2010-Takahara_F-ApJ.715.1248T,Tanaka_SJ-2013-Takahara_F-MNRAS.429.2945T};
\citealp{Murase_K-2016-Kashiyama_K-MNRAS.461.1498M})
\begin{equation}
U_B^{\rm PWN}={B_{\rm PWN}^2 \over {8\pi}}={3\over {4\pi}} \epsilon_B R_{\rm PWN}^{-3}(t) \int_0^t{L_{\rm md}(s)ds}.
\end{equation}
Here $R_{\rm PWN } \sim R_{\min}$ is taken because the deceleration timescale of the ejecta is much larger than the scope of our calculation.
A broken power-law is adopted to describe the energy distribution of leptons in the PWN
\citep{Murase_K-2015-Kashiyama_K-ApJ.805.82M},
 \begin{equation}\label{Eq:ne}
{{d\dot{n}_e}\over{d\gamma_e}} \propto
 \left\{
 \begin{array}{cc}
 \gamma_e^{-q_1},    & \gamma_m \leq \gamma_e < \gamma_b,\\
 \gamma_e^{-q_2},    &  \gamma_b \leq \gamma_e \leq \gamma_M,
 \end{array}
 \right.
 \end{equation}
where $q_1 \sim 1- 2$ ($q_2 \sim 2- 3$) is the low (high)-energy spectral index,
$\gamma_b \sim 10^4 - 10^8$ is the characteristic Lorentz factor of the accelerated leptons in the PWN,
and $\gamma_m$ ($\gamma_M$) is the minimum (maximum) Lorentz factor of leptons.
In this work, we assume $\gamma_m=3$ and $\gamma_M=\sqrt{9m_e^2c^4/(8B_{\rm PWN}q_e^3)}$
\citep{Kumar_P-2012-Hernandez_RA-MNRAS.427L.40K},
where $q_e$ is the charge of leptons, and $m_e$ is the electron mass.

For the synchrotron emission of the PWN, two break frequencies
are related with the leptons' property,
i.e., the characteristic synchrotron frequency $\nu_b$ corresponding to $\gamma_b$,
and the synchrotron cooling frequency $\nu_c$ to
$\gamma_c={6\pi m_e c}/({\sigma_{\rm T} B_{\rm PWN}^2 t})$.
The frequencies are expressed by
\begin{equation}
\nu_i \approx {3\over{4\pi}}\gamma_i^2 {q_e B_{\rm PWN}\over{m_e c}}, \quad i = b,c,m,M,
\end{equation}
where $\sigma_{\rm T}$ is the Thomson cross section (\citealp{Sari_R-1998-Piran_T-ApJ.497L.17S}).
The synchrotron emission is described as follows correspondingly.
In the fast-cooling regime ($\nu_c<\nu_b$), the synchrotron emission flux density $L_{\nu}$
at frequency $\nu$ can be expressed by \citep{Murase_K-2016-Kashiyama_K-MNRAS.461.1498M}
\begin{equation}\label{Eq:PWN_Spectrum_Fast}
\nu L_{\nu}^{^{\rm PWN}}\approx {{\xi L_{\rm md}}\over{2 R_b}}
\left\{
\begin{array}{lcc}
({\nu _c \over \nu _b})^{2 - q_1 \over 2}({\nu_a \over \nu _c})^{3 - q_1\over 2}({\nu \over \nu_a})^{5\over 2}, &\nu_m \leq \nu \leq \nu_a,\\
({\nu _c \over \nu _b})^{2 - q_1 \over 2}({\nu \over \nu _c})^{3 - q_1\over 2}, &\nu_a \leq \nu \leq \nu_c,\\
({\nu \over \nu _b})^{2 - q_1 \over 2},     & \nu_c \leq \nu \leq \nu_b,  \\
({\nu \over \nu _b})^{2 - q_2 \over 2},     & \nu_b \leq \nu \leq \nu_M.
\end{array}
\right.
\end{equation}
In the slow-cooling regime ($\nu_c>\nu_b$),
\begin{equation}\label{Eq:PWN_Spectrum_Slow}
\nu L_{\nu}^{^{\rm PWN}}\approx { {\xi L_{\rm md}}\over{2 R_b}}
\left\{
 \begin{array}{lcc}
({\nu_b\over \nu_c})^{3-q_2\over 2}({\nu_a\over\nu _b})^{3 - q_1\over 2}({\nu \over \nu_a})^{5\over 2}, &\nu_m \leq \nu \leq \nu_a, \\
({\nu_b\over \nu_c})^{3-q_2\over 2}({\nu\over\nu _b})^{3 - q_1\over 2}, &\nu_a \leq \nu \leq \nu_b, \\
({\nu\over \nu_c})^{3-q_2\over 2}, &  \nu_b \leq \nu \leq \nu_c, \\
({\nu\over \nu_c})^{2-q_2\over 2}, &  \nu_c \leq \nu \leq \nu_M.
\end{array}
\right.
\end{equation}
Where $R_b\simeq (2-q_1)^{-1}+(q_2-2)^{-1}$,
the radiation efficiency $\xi=\eta\epsilon_e$ with $\eta=\min\{1,(\nu_b/\nu_c)^{(q_2-2)/2}\}$
(\citealp{Fan_YZ-2006-Piran_T-MNRAS.369.197F}), and $\epsilon_e=1-\epsilon_B$ is adopted.
The critical synchrotron self-absorption (SSA) frequency $\nu_a$ can be calculated with $\tau_a(\nu_a)=1$,
and the SSA optical depth is estimated by (e.g. \citealp{Panaitescu_A-2004-Kumar_P-MNRAS.353.511P};
\citealp{Murase_K-2014-Thompson_TA-MNRAS.440.2528M})
\begin{equation}
\tau_a(\nu) \simeq \xi_{q_1}\frac{q_e n_{\rm ext}R_{\rm PWN}}{B_{\rm PWN} \gamma_m^5}\left({\nu \over \nu_m}\right)^{(-q_1/2-2)},
\end{equation}
where $\xi_{q_1}\simeq 5/3$, and $n_{\rm ext}\simeq n_{\rm PWN}+n_{\rm ej}$ is the number density of leptons.
Here, $n_{\rm PWN}\simeq L_{\rm md}/(4 \pi R_{\rm PWN}^2 \gamma_b m_e c^3)$
and $n_{\rm ej}\simeq 3 M_{\rm ej}/( 4 \pi m_p R_{\rm PWN}^3)$.
Based on Equations~(\ref{Eq:PWN_Spectrum_Fast}) and (\ref{Eq:PWN_Spectrum_Slow}),
one can have $\int_0^{ + \infty } {{L_\nu }d\nu } \approx \eta {\epsilon_e}{L_{{\rm{md}}}}$.
In this work, the effects of the inverse Compton scattering process and the SSA heating are ignored.
The observed flux from a PWN can be expressed as
\begin{equation}
F_\nu ^{{\rm{leak}}}=\frac{{L_\nu}{e^{ - {\tau_{\rm{tot}}}}}}{{4\pi D_L^2}},
\end{equation}
where $\tau_{\rm{tot}}=\sum_1^{N-1} {\Delta {\tau _i}}=\int_{R_{\min}}^{R_{\max}} {\kappa \rho(r) dr}$.
%We assume that the radiation spectra do not change after photons pass through the ejecta.

\section{fitting method}\label{fitting method}
Based on the numerical model developed with section~\ref{basical model},
we jointly fit the observations of AT~2017gfo in UV-optical-IR bands
and the late-time X-ray observations of GRB~170817A.
We implement the Markov Chain Monte Carlo (MCMC) techniques by the use of the Python package {\tt emcee}
\citep{Foreman-Mackey_D-2013-Hogg_DW-PASP.125.306F}.

%It was considered only the UV-optical-IR radiation of AT~2017gfo in \cite{Ren_J-2019-Lin_DB-ApJ.885.60R},
%Here we extend the model to a broader energy band.
The fitting dataset in this work is considered to cover together the AT~2017gfo observations
and X-ray observations of GRB~170817A from 300~days to 1234~days after GW170817 trigger.
The data of AT~2017gfo have been obtained after the extinction correction\footnote{\url {https://kilonova.space/}}.
In addition, the X-ray data are chosen from \cite{Troja_E-2021-O'Connor_B-2021MNRAS.tmp.3213T}
with free photon index $\Gamma$ at the $0.3-10$~keV band.
It is allowed for the possible presence of additional components that may affect the photon index of late observations.

The opacity generated from the bound-free scattering of X-ray photons
is much larger than that for UV-optical-IR photons when the PWN photons crossed the ejecta.
We consider this effect in detail below.
On the one hand, the gray opacity $\kappa_{\rm opt}$ (ignoring the differences of opacity in different bands)
of UV-optical-IR photons is taken as a free parameter, and the optimal result can be obtained by the MCMC method.
On the other hand, when calculating the flux of PWN emission at $0.3-10$~keV band,
the opacity of ejecta is fixed as $\kappa_{\rm X}(E)=\kappa_0(\frac{E}{\rm 1keV})^{-\beta}$,
where $E$ is the energy of X-ray photons,
$\kappa_0=\kappa_{\rm X}(\rm 1keV)=7\times 10^3\;{\rm cm}^2\cdot {\rm g}^{-1}$,
and $\beta = 1$ is assumed (\citealp{Chen_MH-2021-Li_LX-ApJ.919.59C}).
Taking into account the variable opacity, the observed flux from PWN at $0.3-10$~keV band \textbf{is} calculated by
\begin{equation}
\mathcal{F}_{\rm PWN}=\int_{\rm 0.3keV}^{\rm 10keV} F_{\rm X}^{\rm{leak}}(E)dE.
\end{equation}
Here
\begin{equation}
F_{\rm X}^{\rm{leak}}(E)=\frac{L_{\rm X}(E)e^{-\tau_{\rm tot,X}(E)}}{4\pi D_L^2}
\end{equation}
is the flux density of leakage after X-ray photons cross the ejecta,
where the total optical depth from the ejecta of X-ray photons with energy $E$ is substituted as
$\tau_{\rm{tot,X}}(E)=\int_{R_{\min}}^{R_{\max}} {\kappa_{\rm X}(E) \rho(r) dr}$.

Except for the PWN radiation, the external shock emission of the jet have still affect the observations to the remnant of GW170817 event.
However, the post-break light curves of GRB~170817A afterglow in different jet models are different
based on, for example, the specific jet profile, the consideration of the lateral expansion, and the selection of parameters
(e.g., \citealp{Ren_J-2020-Lin_DB-ApJ.901L..26R,Lamb_GP-2020-Levan_AJ-ApJ.899..105L,
Li_L-2021-Dai_ZG-ApJ.918.52L,Hajela_A-2021-Margutti_R-arXiv210402070H,
Troja_E-2021-O'Connor_B-2021MNRAS.tmp.3213T}).
It is difficult to proceed with the work if all the models at once are taken into account.
Therefore, we choose a phenomenological formula to describe the flux at  $0.3-10$~keV band of jet emissions,
\begin{equation}
\mathcal{F}_{\rm afterglow}=120\cdot t^{-p}\; {\rm erg\cdot cm^{-2}\cdot s^{-1}},
\end{equation}
where $t$ ($> 200$~days) is the observed time, and $p=2.16$ is the spectrum index of electrons in the GRB jet.
Here $p$ is chosen from the fits of the early spectra of GRB~170817A afterglow.
Thus, the observed total flux $\mathcal{F}_{\rm total}$ at $0.3-10$~keV band should be considered by
\begin{equation}
\mathcal{F}_{\rm total}=\mathcal{F}_{\rm PWN} + \mathcal{F}_{\rm afterglow}.
\end{equation}
%The curve of $\mathcal{F}_{\rm afterglow}$ is shown in Figure~{\ref{MyFig2}}.

In \cite{Ren_J-2019-Lin_DB-ApJ.885.60R},
the value of the characteristic Lorentz factor of the accelerated leptons in the PWN is fixed to $\gamma_b=10^4$,
and the fraction of the magnetic energy density to the total energy density is fixed to be $\epsilon_B=10^{-2}$.
Both of them are taken to be free parameters in this work.
Overall, the eleven physical parameters in our models, \{$q_1,q_2,\gamma_b,\epsilon_B,L_{\rm md,0},t_{\rm sd},M_{\rm ej},\kappa_{\rm opt},
v_{\min},v_{\max},\delta$\}, are fitted as free parameters.

\section{Fitting Result and Discussion}\label{discussion}
\begin{figure}
\centering
\includegraphics[width=0.45\hsize]{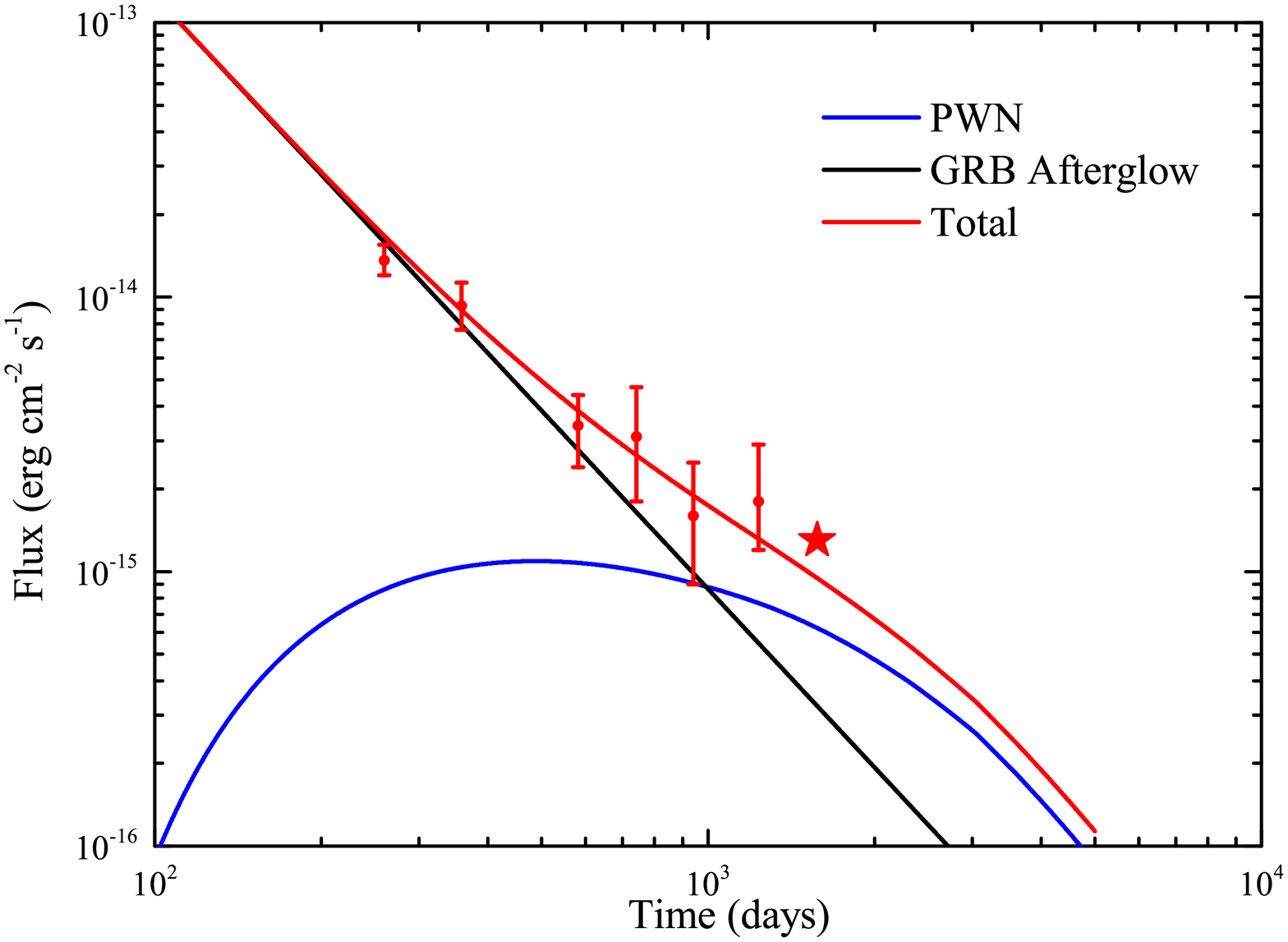}
\includegraphics[width=0.45\hsize]{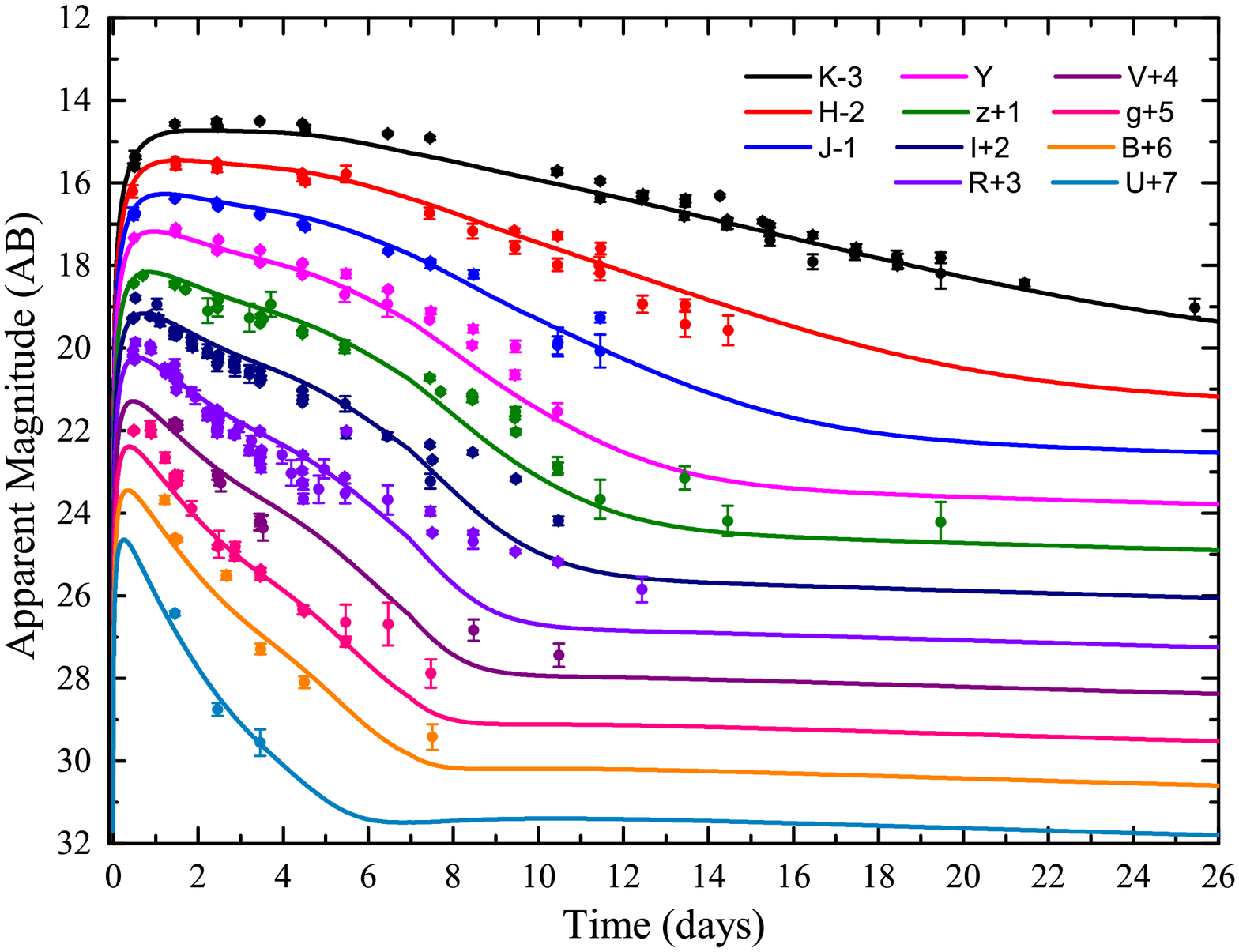}
\caption{Left Panel: The X-ray light curve of GRB~170817A afterglow with the addition of PWN emission.
Data are taken from the free-$\Gamma$ column, $0.3-10$~keV flux,
in Table~1 of \citet{Troja_E-2021-O'Connor_B-2021MNRAS.tmp.3213T}.
We also mark the new observation at $\sim 1575$~days with a star, as given by
\citet{Hajela_A-2021-Margutti_R-GCN.31231....1H},
where the photon index $\Gamma=1.6$ was fixed but the $1\sigma$ uncertainties were not reported. We notice the data is consistent with our model.
Right Panel: The multiband light curves of AT~2017gfo fitted by our model.
The data are taken from \url{https://kilonova.space/} and the extinction correction is performed.}
\label{MyFig2}
\end{figure}

The model parameters at the $1\sigma$ confidence level are given in Table~{\ref{MyTabA}},
where the projections of the posterior distribution for the parameters are presented in Figure~{\ref{MyFig1}}.
As shown in Figure~{\ref{MyFig2}}, the joint fitting shows a good result for the AT~2017gfo observations, and
the additional X-ray component formed by the PWN could explain the flattening of the X-ray observations of GRB~170817A.

\subsection{Comparison with previous fitting}
\begin{figure}
 \centering
 \includegraphics[width=0.8\hsize]{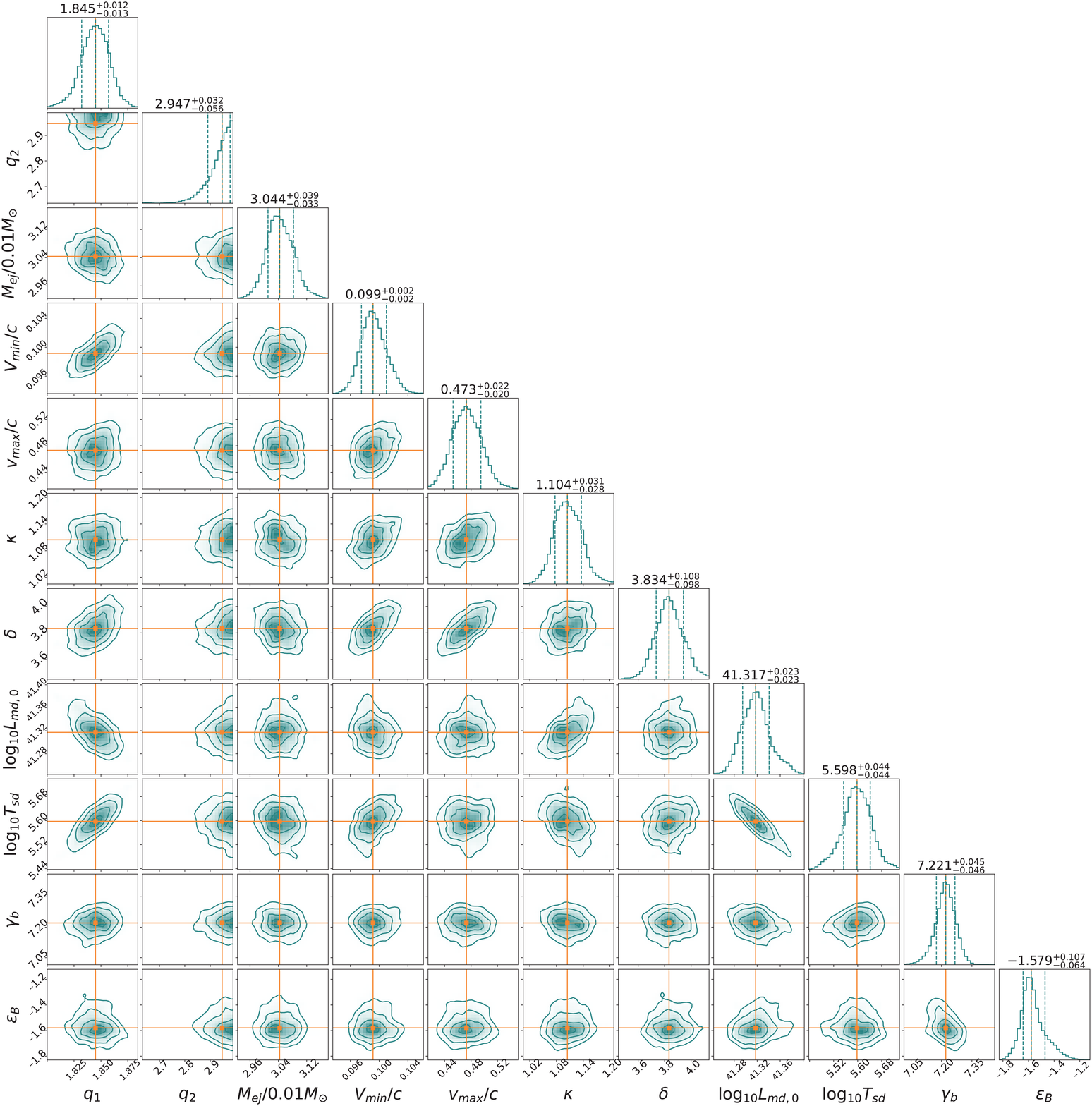}
 \caption{Posterior probability density contours for the physical parameters from the MCMC sampling.}
 \label{MyFig1}
\end{figure}

Comparing the results in this paper and \cite{Ren_J-2019-Lin_DB-ApJ.885.60R} in Table~{\ref{MyTabA}},
we find that the optimal values of most of the parameters change slightly.
Nonetheless, the spin-down timescale $t_{\rm sd}$, remains the most important parameter,
changing from $2.34\times 10^5$~s to $3.96\times 10^5$~s.
This change is based on the enlargement of the dataset caused by the addition of X-ray data.
Using Equation~{\ref{tsdgw}}, the newly fitting result of the ellipticity of the central NS can be obtained,
$\epsilon=1.52\times 10^{-4}I_{45}^{-1/2}P_{0,-3}^2$, which is slightly less than that of  \cite{Ren_J-2019-Lin_DB-ApJ.885.60R}.
Meanwhile, the initial spin-down luminosity $L_{\rm md,0}$ is the same as before,
which means the early-time kilonova observations could well estimate the spin-down luminosity of central NS.
However, the degeneracy between $t_{\rm sd}$ and $L_{\rm md,0}$ also indicates the possibility of the other parameter combinations, as shown in Figure~{\ref{MyFig1}}. But we note it does not affect the conclusion here.

\begin{figure}
\centering
\includegraphics[width=0.8\hsize]{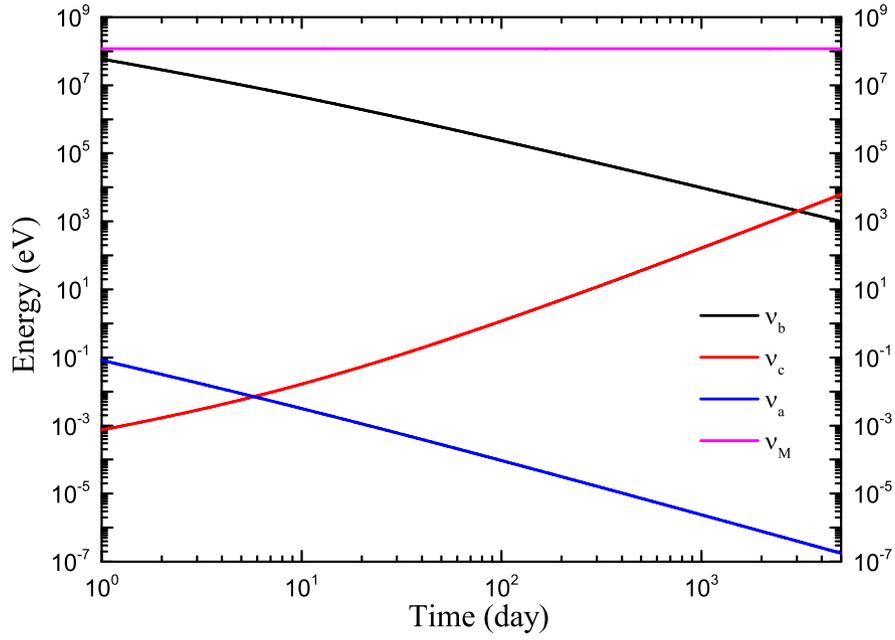}
\caption{The evolution of $\nu_b$, $\nu_c$, $\nu_a$, and $\nu_M$ of the PWN
using the MCMC optimal results.
The picture shows that the PWN is in the fast-cooling regime in $\sim 3000$~days
after the merger. The transition of $\nu_b$ and $\nu_c$ in the X-ray band will lead to
softening of the X-ray spectrum.}
\label{MyFig3}
\end{figure}

There is an important change in the parameters of the spectral energy distribution in the PWN.
We have gotten the optimal results of $q_1=1.85$, $\gamma_b=1.66\times10^7$, and $q_2\sim 3$ in this work.
Since, in the early fast-cooling stage, the frequencies of most of the PWN photons
are located within the range of  $\nu_c \leq \nu \leq \nu_b$,  the fitting of $q_1$ is
determined by the observations of AT~2017gfo, and its changeless result is easy to understand
(see Equation~{\ref{Eq:PWN_Spectrum_Fast}}).
We show the evolution of
$\nu_b$, $\nu_c$, $\nu_a$, and $\nu_M$ in Figure~{\ref{MyFig3}} for the MCMC results.
As shown in Figure~{\ref{MyFig3}},
lines of $\nu_b$ and $\nu_c$ cross each other at $\sim3000$~days after the merger,
and the conversion frequency is located in the X-ray band.
One can find that the PWN is still in the fast cooling regime in $\sim3000$ days after the merger.
It means that the constraints on $q_2$ from observations are very weak.
This is why $q_2$ is not well constrained, as shown in Figure~{\ref{MyFig1}}.
Based on Equation~{\ref{Eq:PWN_Spectrum_Fast}},
the photon index of PWN photons is $\Gamma=1+q_1/2=1.92$,
suggests a soft spectrum as seen in the X-ray observations \citep{Troja_E-2021-O'Connor_B-2021MNRAS.tmp.3213T}.

%Additionally, we have gotten the optimal results of
%$\gamma_b=1.66\times10^7$ and $q_2\sim 3$ in this work.
%Although these values are much larger than those in \cite{Ren_J-2019-Lin_DB-ApJ.885.60R},
%they \textbf{are consistent with} typical values of the parameters of PWNe observed in our galaxy
%(e.g., \citealp{Zhu_BT-2018-Zhang_L-A&A.609A.110Z}).

\subsection{The trend of X-ray observations}
The X-ray afterglow of GRB~170817A is predicted in this work
to evolve from flattening to a steep decline, as shown in the left panel of Figure~{\ref{MyFig2}}.
In terms of the trend of X-ray observation,
the kilonova ejecta-PWN system model and the BH fallback accretion model
\citep{Ishizaki_W-2021-Ioka_K-ApJ.916L.13I,Ishizaki_W-2021-Kiuchi_K-ApJ.922.185I,
Metzger_BD-2021-Fernandez_R-ApJ.916L.3M} have similar predictions.
Nevertheless, there are certain differences in the slope of decay predicted by the two models.
The kilonova ejecta-PWN system model has a steeper decay index than
that in the BH fallback accretion model, i.e., steeper than $t^{-5/3}$.
In addition, the BH fallback accretion model predicts
an approximate blackbody spectrum of the additional component.
However, it is a power-law spectrum for the additional PWN component.
Future observations will test our model.

What we should note is that the influence of
one or more extra sources on X-ray observations is possible.
The models of the kilonova afterglow
\citep{Hajela_A-2021-Margutti_R-arXiv210402070H,Nedora_V-2021-Radice_D-MNRAS.506.5908N},
and even the count-jet \citep{Li_LB-2019-Geng_JJ-ApJ.880...39L,Troja_E-2021-O'Connor_B-2021MNRAS.tmp.3213T}
are predicted to have an impact during this stage.
The combined effect could complicate the evolution of future X-ray observations.
To distinguish these components,
continuous observations at the radio band is an important approach
(see Section~{\ref{Conclusion}}).

\subsection{The radio emission from PWN}
\begin{figure}
\centering
\includegraphics[width=0.8\hsize]{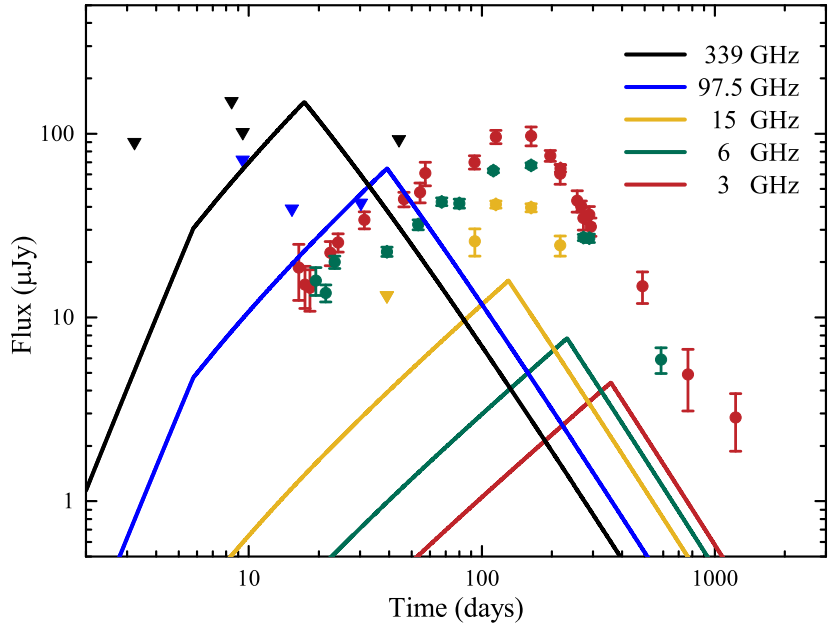}
\caption{Multiband radio light curves emitted from the kilonova ejecta-PWN system of the GW170817 event remnant are
calculated by the optimal parameters estimated from the MCMC sampling.
The same bands are represented by the same colors correspondingly,
where the observational data are described with circles,
and the triangles are the upper limits.
The data are chosen from \citet{Makhathini_S-2021-Mooley_KP-ApJ.922.154M} and
\citet{Balasubramanian_A-2021-Corsi_A-ApJ.914L..20B}.}
\label{MyFig4}
\end{figure}

We also check the light curves on radio bands emitted by the PWN.
The model curves at radio bands shown in Figure~{\ref{MyFig4}} are
calculated with the MCMC fitting optimal values of the parameters.
We find that the PWN's radio emission flux is below the observations,
which verifies the dominance of the GRB afterglow to the observations.
Because the late-time radio emission from the PWN is significantly
dimmer than the emission from external forward shock of the jet,
it is not expected to see an effect of the PWN in future observations.

More signals of binary NS mergers will be detected by GW detectors in near future.
Due to the Doppler effect of the large off-axis viewing angle of the merger jet,
the prompt GRB radiation is hard to detect for most parts of events.
Meanwhile, the afterglows of merger jets are also going to be much dimmer.
As shown by our results, however, if a long-lived NS is formed during the binary NS merger,
the kilonova ejecta-PWN system could power a long-lasting radio emission.
This trait considerably increases the probability of detecting EM counterparts to GW events.

\section{Conclusions}\label{Conclusion}
In this paper, we have studied the broadband radiation behavior of the kilonova ejecta-PWN system,
and fitted jointly the observations of AT~2017gfo in UV-optical-IR bands
and the late-time X-ray observations of GRB~170817A.
The kilonova ejecta-PWN system could explain the observations of AT~2017gfo based on two effects,
namely the leakage of PWN radiation and its heating influence on the merger ejecta.
An additional component shows a signature in the late-time X-ray observations of GRB~170817A
when the ejecta is optically thin to X-ray photons from the PWN.
We have shown that the trend of future X-ray observations will be dominated by the PWN emission,
from flattening to a steep decline, until other sources have their impact, e.g., the kilonova afterglow.

The constrained ranges of the parameters of the central NS in this work, i.e.,
the dipole magnetic field strength and the ellipticity,
are almost the same as those in the previous works
\citep{Yu_YW-2018-Liu_LD-ApJ.861..114Y,Ren_J-2019-Lin_DB-ApJ.885.60R}
and are consistent with the limits set by \cite{Ai_SK-2020-Gao-H-ApJ.893..146A}.
It is suggested that a long-lived NS still exists in the merger remnant center a few years after the GW170817 event.
Additionally, the new fitting result of the characteristic Lorentz factor
of the accelerated leptons indicates that the NS wind is highly magnetized.
Note that the parameter values of the PWN are in a reasonable range
compared with the PWN observations in our galaxy.

We also showed the light curves of radio emission from the PWN
obtained by the optimal values of the fitting parameters.
We found that the flux of radio emission from the PWN is lower than the observations of GW170817 event counterpart.
This indicates that the radio observations are dominated by the GRB jet afterglow,
and not affected by PWN radiation even in the future.
This leads to the difference between the kilonova ejecta-PWN system model and the kilonova afterglow model.
Besides, the kilonova ejecta-PWN system model and the fallback accretion model
have similar predictions of the trend of X-ray observation but certain differences
in the slope of decay predicted by the two models. A difference is
that the decay index in the kilonova ejecta-PWN system model is steeper than that (e.g, $-5/3$)
in the fallback accretion model.
In addition, an approximate blackbody spectrum of the additional component is suggested by the fallback accretion model
but a power-law spectrum is predicted in the kilonova ejecta-PWN system model.
Future continuous observations will test our model.

What should be pointed out is that the radio and X-ray emission from the kilonova ejecta-PWN system model
could be an important probe of binary NS mergers from days to years after the GW signals.
There is a lack of jet-related signals to trigger the detectors when the axis of a merger-produced jet
is far away from the line of sight.
Differently, the quasi-isotropic, long-existing emission of a kilonova ejecta-PWN system gives rise to more advantage in observations.
Its flux is related to the luminosity and total energy released by the central engine.
This means that the radiation behavior of a kilonova ejecta-PWN system
would directly provide the information of the merger remnant itself,
possibly revealing the properties of a newborn NS and an accompanying young PWN.

\section*{acknowledgments}
RJ would like to thank Ken Chen, Da-Bin Lin and Yun-Wei Yu for helpful discussions.
We thank the referee for his scrutiny,
kind comments and helpful suggestions to help improve this article.
This work was supported by the National Key Research and Development Program of China (grant No. 2017YFA0402600),
the National SKA Program of China (grant No. 2020SKA0120300),
and the National Natural Science Foundation of China (grant No. 11833003).
%We thank the anonymous referee, for the scrutiny and useful comments and suggestions.

\section*{data availability}
The data underlying this paper will be shared on reasonable request to the corresponding authors.
%%%%%%%%%%%%%%%%%%%%%%%%%%%%%%%%%%%%%%%%%%%%%%%%%%%%%%%%%%%%%%%%%%%%%%%%%%%%%%%%%%%%%%%%%%%%%%%%%%%
%%%%%%%%%%%%%%%%%%%%%%%%%%%%%%%%%%%%%%%%%%%%%%%%%%%%%%%%%%%%%%%%%%%%%%%%%%%%%%%%%%%%%%%%%%%%%%%%%%%

%\bibliographystyle{mnras}
%\bibliography{bibliography}

\onecolumn
\begin{table}
\centering
\caption{Parameters estimated from the MCMC sampling.}
\begin{threeparttable}[b]
\label{MyTabA}
\begin{tabular}{ccc||c}
\hline \hline
Parameter   & Constraint    & Range\tnote{1}  & Pervious      \\ \hline
 ${\rm log}_{10}L_{\rm md,0}$~($\rm{erg\cdot s}^{-1}$)                          &$41.317_{-0.023}^{+0.023}$       & [39, 43]         &$1.85_{-0.23}^{+0.85}\times 10^{41}$  \\
 ${\rm log}_{10}t_{\rm sd}$~(s)                                                                       & $5.598_{-0.044}^{+0.044}$       & [3, 7]                &$2.34_{-0.94}^{+0.57}\times 10^{5}$  \\
 $M_{\rm{ej}}/0.01M_{\odot}$                                                                         & $3.044_{-0.033}^{+0.039}$         & [1, 10]            &$3.52_{-0.11}^{+0.06}$\\
 $\kappa_{\rm opt}$~($\rm{cm}^2\cdot \rm{g}^{-1}$)                            & $1.104_{-0.028}^{+0.031}$         & [0.1, 10]        &$1.69_{-0.09}^{+0.06}$  \\
 $v_{\rm{min}}/c$                                                                                              & $0.099_{-0.002}^{+0.002}$         & [0.05, 0.15]   &$0.10_{-0.00}^{+0.01}$ \\
 $v_{\rm{max}}/c$                                                                                             & $0.473_{-0.020}^{+0.022}$         & [0.3, 0.6]         &$0.34_{-0.01}^{+0.01}$ \\
 $\delta$                                                                                                              & $3.834_{-0.098}^{+0.108}$          & [1, 5]              & $2.47_{-0.32}^{+0.02}$ \\
 $q_1$                                                                                                                  &$1.845_{-0.013}^{+0.012}$            & [1, 2]              &$1.83_{-0.05}^{+0.02}$ \\
 $q_2$                                                                                                                  & $2.947_{-0.056}^{+0.032}$           & [2, 3]             & $2.25_{-0.05}^{+0.10}$ \\
 ${\rm log}_{10}{\gamma_b}$                                                                       & $7.221_{-0.046}^{+0.045}$           & [4, 8]             & $10^4$  \\
 ${\rm log}_{10}{\epsilon_B}$                                                                       &$-1.579_{-0.064}^{+0.107}$           &[-3,-0.5]        &$0.01$  \\
 $\kappa_{\rm X}({\rm 1keV})$~($\rm{cm}^2\cdot \rm{g}^{-1}$)      & $7\times10^3$                                            & --                     & -- \\
 $\alpha$                                                                                                           & 1                                                       & --                     &  1 \\
\hline \hline
\end{tabular}%
\begin{tablenotes}
\item[1] Priors are uniformly distributed.
\end{tablenotes}
\end{threeparttable}
\par
\end{table}

\clearpage

\clearpage

\clearpage

\clearpage

\bsp	% typesetting comment
% Don't change these lines
\label{lastpage}
\end{document}